\documentclass[11pt]{article}
\usepackage{amsfonts}
\usepackage{amssymb}
\usepackage{amsmath}
\usepackage{indentfirst}
\usepackage[latin1]{inputenc}
\usepackage[dvips]{graphicx,color}
\usepackage{fullpage}

\newcommand{\bea}{\begin{eqnarray}}
\newcommand{\eea}{\end{eqnarray}}

\def\com#1#2{\Big[#1,#2\Big ]}

\def\be{\begin{equation}}
\def\ee{\end{equation}}

\def\fr{\frac}
\def\a{\alpha}

\def\e{\epsilon}

\def\l{\lambda}
\def\m{\mu}
\def\n{\nu}
\def\n{\nu}

\def\s{\sigma}

\def\nn{\noindent}
\def\no{\nonumber}

\def\uq{U_q(su(2))}
\def\half{\frac{1}{2}}
\newcommand{\Hil}{\mathcal{H}}

\title{Index Theorem for the $q$-Deformed Fuzzy Sphere}
\author{E. Harikumar$^a$\footnote{harisp@uohyd.ernet.in}, Amilcar R. Queiroz$^b$\footnote{amilcar@fma.if.usp.br}  and P. Teotonio-Sobrinho$^b$\footnote{teotonio@fma.if.usp.br} \\ \\
  $^a$ School of Physics, University of Hyderabad, \\ Central University P O,
  Hyderabad-46, AP, India, Pin 500 046 \\ \\ $^b$ Instituto de F\'{\i}sica,
  Universidade de S\~{a}o Paulo,\\ Caixa Postal 66318, 05315-970, S\~ao
  Paulo, SP, Brazil}

\begin{document}
\maketitle

\begin{abstract}
  We calculate the index of the Dirac operator defined on the q-deformed
  fuzzy sphere. The index of the Dirac operator is related to its net
  chiral zero modes and thus to the trace of the chirality operator.  We
  show that for the q-deformed fuzzy sphere, a $\uq$ invariant trace of
  the chirality operator gives the q-dimension of the eigenspace of the
  zero modes of the Dirac operator. We also show that this q-dimension is
  related to the topological index of the spinorial field  
  as well as to the fuzzy cut-off parameter. We then
  introduce a q-deformed chirality operator and show that its $\uq$
  invariant trace gives the topological invariant index of the Dirac
  operator. We also explain the construction and important role of the
  trace operation which is invariant under the $\uq$, which is the symmetry
  algebra of the q-deformed fuzzy sphere. We briefly discuss chiral
  symmetry of the spinorial action on q-deformed fuzzy sphere and the
  possible role of this deformed chiral operator in its evaluation using
  path integral methods.
\end{abstract} 
\newpage
\section{Introduction}

The index of Dirac operator is known to be related to topological
invariance of the underlying space where it is defined. The index of a
self-adjoint operator ${\cal A}$ is defined 
as the difference between the number of the zero modes of ${\cal A}$ and its
adjoint ${\cal A}^\dagger$\cite{topicsing}. Thus, in the case of Dirac
operator, the index is a measure of the net chiral zero modes and thus
it is related to the chirality operator $\Gamma$ and also to the
chiral symmetry of the fermionic field theories. The study of
index theory of the Dirac operator have been of interest for the
understanding of chiral anomaly apart from its importance in
understanding the topological features of the underlying space. For
fermions coupled to gauge fields, it is known that the index can be
expressed in terms of the field strength of gauge fields alone. The
index is exactly the Pontryagin index which is a
topological invariant quantity of the underlying
space\cite{topicsing}. In the Fujikawa mechanism \cite{fujikawa}, this
topological term is added to the classical action so as to make the
the partition function invariant under the chiral transformations
which is a symmetry of the classical theory \cite{topicsing, fujikawa, peshkin}.

The study of the index theorem for Dirac operators on different spaces are
of interest not only for the investigation of the topological features but
also for the understanding of chiral symmetry. This relation between the
zero modes, topological charge of the underlying space and chiral anomaly
have been analysed for Dirac operators on various spaces such as two
sphere and fuzzy sphere\cite{pp, bal, watta, kullish, grose}.

In this paper we obtain the index theorem for a recently constructed
\cite{us} Dirac operator on the q-deformed fuzzy sphere. An important
feature of this $D_q$ is its invariance under the action of $\uq$. Such
symmetry plays a determinant role in the computation of the spectrum of
$D_q$. The eigenfunctions of $D_q$ obtained in \cite{us} can be classified
into $\pm$ chiral subspaces using the chirality operator $\Gamma$.  We
show a possible modification of the chirality operator obtained in
\cite{us} and this deformed chiral operator $\tilde\Gamma$ also splits the
spinor space into $\pm$ chiral subspace and reduces to chiral operator on
fuzzy sphere in the limit $q\to 1$.  Using relation of the index of $D_q$
to the trace of the chirality operator, we explicitly find the index of
$D_q$. In obtaining this, we show the important role played by the $\uq$
invariant trace. We show that the trace of the chirality operator gives a
relation between $q$-dimension of the eigenspace of the zero modes of
$D_q$ and the topological index of the spinor field.  We then calculate
the index using the deformed chirality operator ${\tilde\Gamma}$. We
observe that the indices obtained from chirality operator $\Gamma$ as well
as deformed chirality operator $\tilde\Gamma$, have the proper limit with $q\to 1$.

The relation between Hopf algebra (also known as quantum groups) and
noncommutative spaces is a well estabilished matter. In particular, the
geometrical and topological aspects of these spaces have been studied in
\cite{landi, landi2}. The construction of field theories on noncommutative
geometries with underlying symmetry being $\uq$ or its quantum dual
$SU(2)_q$ has been explored in \cite{steinacker}.

The study of field theory models on non-commutative space in
general and on fuzzy $S^2$ and $S_{qF}^2$ in particular, are of
importance as they provide alternate, finite dimensional, regularised
models. These models may have advantages such as the absence of
fermion doubling problem which plague the usual lattice
formulation\cite{bty}.  Attempts to construct gauge field theories on
such manifolds have also been undertaken \cite{hs}.  We hope that the
understanding of the index theory and chiral invariance in these
models ( à la Fujikawa mechanism ) can be helpful in providing us
valuable hints in the construction of gauge theories on $S_{qF}^2$.

The $q$-deformed fuzzy sphere $S^{2}_{qF}(N)$ is described by a finite
dimensional matrix algebra and carries an action of $\uq$. The number
$N$ which fixes the dimension of the matrices is called the fuzzy
cut-off parameter. In the limit $q\to 1$, this reduces to the usual
fuzzy sphere defined as the matrix algebra invariant under $su(2)$.

In other words, $\uq$ is the symmetry of the q-deformed fuzzy sphere
just as $su(2)$ is the symmetry of the usual fuzzy sphere.
We can describe  $\uq$ as the algebra generated by the operators $J_{\pm}$ and $K\equiv
q^{J_3}$, where $q$ can be a positive real number or a root of
unity (i.e., $q=e^{\frac{2 \pi i}{p}}$ with $p$ being a positive
number) such that the operators satisfy the relations
\be
  \label{uq}
  \com{J_+}{J_-}=\frac{K-K^{-1}}{q^{\half}-q^{-\half}}~~~ \textrm{and} ~~~
  KJ_{\pm}K^{-1}=q^{\pm 1}J_{\pm}.
\ee
Here we used the definition of $q$-number of $x$ 
\begin{equation}
  \label{qnodef}
  [x]\equiv [x]_q =\frac{q^{\frac{x}{2}}-q^{-\frac{x}{2}}}{q^{\frac{1}{2}}-q^{-\frac{1}{2}}} .
\end{equation}

$\uq$ being a Hopf algebra
\cite{qgroups}, the Hopf algebra structures, namely, 
co-product~$\Delta$, antipode~$S$ and co-unit~$\e$ of $\uq$  are given by
\begin{eqnarray}
  \label{eq:q-hopf-structures}
 & \Delta(J_{\pm})= J_{\pm}\otimes K^{\half}+K^{-\half}\otimes J_{\pm},~~~
  \Delta(K)=K\otimes K& \\
  &S(J_{\pm})=-K^{\half}J_{\pm}K^{-\half}, ~ ~ ~
  S(K)=K^{-1}& \\
  &\epsilon(J_{\pm})=0,~~~ \epsilon(K)=\epsilon(\mathbb{I})=1&
  \end{eqnarray}
where $[n]_q!=[n]_q[n-1]_q...[1]_q$. Observe that the Hopf algebra structures
are not preserved by $q\leftrightarrow q^{-1}$. We, thus, have two Hopf
algebras $\uq$ and $U_{q^{-1}}(su(2))$.

After briefly stating the well known index theorem in a language suitable
for our purpose in the next section, we summarise the essential details of
the constructions of spinor module and its chiral decomposition, Dirac
operator and its spectrum. This is done first for the fuzzy sphere in
section 3, where we also show the validity of index theorem and then in
section 4 for the q-deformed fuzzy sphere with more details about the
$\uq$-invariant trace, which will be crucial for our later discussions. In
section 5 we present our main result, namely the derivation of the index
theorem for the Dirac operator $D_q$ on $S_{qF}^2$. Here we show that the
trace of the chirality operator $\Gamma$ relates the q-dimension of the
eigenspace of $D_q$ to the topological index of the spinor field and also to the fuzzy cut-off. We then
define a $q$-deformed chirality operator $\tilde\Gamma$ and show that its
trace gives the topological invariant index of $D_q$. We conclude in
section 6 with comments about the important role of $\uq$-invariant trace
and its relevance to the chiral anomaly of the theory defined on
$S_{qF}^2$.

\section{Index Theorem }

The index of an operator $D$ is defined as the difference between the
number of the zero modes of D and that of $D^\dagger$, i.e., 
\begin{equation} 
\label{index-dimension}
Index~ D=
dim (Ker D) - dim (Ker D^\dagger)
\end{equation}
 where the $dim (Ker) $ is the
dimension of the space of zero modes of the respective operators.  The
index theorem relates the topological invariance of the manifold where the
operator is defined to its index\cite{topicsing}. In the case of the Dirac
operator in Euclidean space, the index $\n$ is the difference between
$n_+$, the number of zero modes with positive chirality and $n_-$, the
number of zero modes with negative chirality.  It is a well known fact
that the index of the Dirac operator is a topological invariant, i.e., \be
n_+-n_-=\nu=\fr{1}{32\pi^2}\int d^4x~\e_{\m\n\l\s}F^{\m\n}F^{\l\s}.  \ee
In order to evaluate the index of the Dirac operator, it is sufficient to find
all the zero modes and classify them according to chirality. Thus we see
that the index of the Dirac operator is the same as the trace of the
chirality operator $\Gamma$ restricted to the space of zero modes of the
Dirac operator. Here, to clarify the evaluation of the index, we start
with a generic self-adjoint operator
\begin{enumerate}
\item $D$ which act on a vector space $V$,
\item An involution operator $\Gamma$ on $V$( i.e., 
  $\Gamma^2=1$) which anti-commutes with $D$, i.e., $D\Gamma+\Gamma D=0$.
\end{enumerate}
Once these conditions are fulfilled, one can see that
\be 
Tr ~\Gamma= |n_+ -n_-|
\label{tr}
\ee
where the trace is evaluated on $V$ and $n_\pm$ are the number of
$\pm$-chiral zero modes. We can simplify the
calculation of the above trace by splitting $V$ into two subspaces with respect to the
eigenvalues of $D$. Thus let $V_0$ be the subspace of $V$
such that $D(V_0)=0$ and $V_1$ the subspace of $V$ such
that $D(V_1) \neq 0$, then $V=V_0\oplus V_1$. The trace can be
written as a sum of two traces, one of them taken on $V_0$ and the
other on $V_1$, i.e.,
\begin{equation}
  \label{trace-splitted}
  Tr(\Gamma)=Tr_{\scriptscriptstyle D=0}(\Gamma)+Tr_{\scriptscriptstyle D\neq 0}(\Gamma).
\end{equation}
One can easily see that the second term on the right hand side vanishes. Indeed, if we consider
the operator $|D|=(DD^{\dagger})^{\half}>0$, then 
\begin{eqnarray*}
  Tr_{\scriptscriptstyle D\neq 0} (\Gamma)&=&Tr_{\scriptscriptstyle D\neq 0}(\frac{D}{|D|} \Gamma
  \frac{D^{\dagger}}{|D|}) \\
              &=& - Tr_{\scriptscriptstyle D\neq 0}(\frac{D}{|D|} \frac{D^{\dagger}}{|D|}
              \Gamma)=-Tr_{\scriptscriptstyle D\neq 0}(\Gamma),
\end{eqnarray*}
where we have used the facts $D^{\dagger}=D$, $|D|\Gamma=\Gamma |D|$ and
$\Gamma D+D \Gamma=0$. Thus we get
\be
  \label{eq:index-theorem-for-D} 
Tr (\Gamma) = Tr_{\scriptscriptstyle D=0}(\Gamma)
=Tr_{\scriptscriptstyle D=0}(\frac{1+\Gamma}{2}-\frac{1-\Gamma}{2})=n_+-n_-,
\ee
where $Tr_{\scriptscriptstyle D=0}$ is the usual trace on $V$
restricted to the subspace $V_0$ of the zero modes of the operator
$D$. The number $n_+$ is the dimension of the subspace of $V_0$ with
positive chirality ($\Gamma V_0^+=V_0^+$), and likewise $n_-$ is the
dimension of the subspace of $V_0$ with negative chirality ($\Gamma
V_0^-=-V_0^-$). Note that the operators $(1\pm \Gamma)/2$ project the spinors to $\pm$ chiral subspaces and in particular on
$V_0^{\pm}$ respectively, in this case.

\section{Dirac Operator on the Fuzzy Sphere}

In this section, we briefly recall the essential features of this
Dirac operator and its spectrum on fuzzy sphere \cite{gkp} and show
the validity of the index theorem.

The Dirac operator on the fuzzy sphere maps the $su(2)$
spinor module
($\mathcal{S}^J_{k}$) to itself. The spinorial field belonging to this
spinor module $\mathcal{S}^J_{k}$ is defined as
\begin{equation}
  \label{eq:spinorial-fields}
  \Psi=\Psi^+(a^{\dagger}_i,a_i,b^{\dagger},b)+\Psi^-(a^{\dagger}_i,a_i,b^{\dagger},b)=f(a^{\dagger}_i,a_i)b+g(a^{\dagger}_i,a_i)b^{\dagger},
\end{equation}
where $a_i^{\dagger},a_i, i=1, 2$ are two sets of bosonic creation and
annihilation operators, $b^{\dagger},b$ is a set
of fermionic creation and annihilation operators ($\{b,b^{\dagger}\}=1$),
and the functions $\Psi^{\pm}$ can be written as a linear combination
of monomials with fixed topological index $2k\in {\mathbb{Z}}$. Thus we have
\begin{equation}
  \label{eq:spinorial-field-bosonic-part}
  \Psi^{\pm}=\sum_{m_1,m_2,\mu, n_1, n_2,\nu}c_{m_1,m_2,\mu,n_1,n_2,\nu}a_1^{\dagger \ m_1}a_2^{\dagger \
    m_2}b^{\dagger\mu}a^{n_1}_1a^{n_2}_2b^{\nu}
\end{equation}
where $m_1,m_2, n_1, n_2$ are non-negative integers and $\mu,\nu=0,1,
\mu+\nu =1$ satisfying $m_1+m_2+\mu\le M$,
$n_1+n_2+\nu\le N$, $M-N=2k$ apart from the condition
$m_1+m_2+\mu-n_1-n_2-\nu=2k$. We also have $M+N=2J$ which is the fuzzy
cut-off parameter. Some properties we have to bear in mind concerning these
spinorials $\Psi^\pm$ are
\begin{itemize}
\item they map the Fock space $\mathcal{F}^{\mu}_N$ to the Fock Space
  $\mathcal{F}^{\nu}_M$. These Fock spaces are Hilbert spaces of
  representations of $su(2)$ and are given by 
  \begin{equation}
    \label{eq:fock-space}
    |n_1,n_2;\nu \rangle=\frac{1}{\sqrt{n_1!n_2!}} a_1^{\dagger
      {n_1}}a_2^{\dagger {n_2}}b^{\dagger\nu}
|0\rangle,~~{n_1}+{n_2}+\n=N.
  \end{equation}
\item The bosonic parts of the spinor $\Psi$, i.e., $f$ and $g$ in (\ref{eq:spinorial-fields}), are
  tensor representations with even dimensions of $U(su(2))$. Therefore,
  the angular momentum number $l$ of these operators are half-integers
  $\frac{2s+1}{2}$, where $s\in\mathbb{Z}$. Further, the bosonic operator
  $f$ can be written as a $\frac{M}{2}\otimes \frac{(N-1)}{2}$ matrix,
  and the bosonic operator $g$ can be written as a $\frac{(M-1)}{2}\otimes
  \frac{N}{2}$ acting on the column vector representation of the Fock
  space (\ref{eq:fock-space}). Thus $f$ and $g$ can be written in
  terms of the tensor operators belonging to the half-integer
  representations
\bea
\fr{M}{2}\otimes\fr{N-1}{2}=|k+\half|\oplus......\oplus(J-\half)\label{dsum1}\\
{\rm ~and~}~~~~ \fr{M-1}{2}\otimes
\fr{N}{2}=|k-\half|\oplus......\oplus(J-\half) 
\label{dsum2}
\eea 
respectively where
$J=\fr{M+N}{2}$.
\item  On the spinorial module $\mathcal{S}_k^{J}$ we define the chirality operator as
\begin{equation}
  \label{eq:chirality-operator}
  \Gamma \Psi = -[b^{\dagger}b,\Psi],
\end{equation}
which has $\pm 1$ eigenvalues. We denote the respective eigenspaces as $\mathcal{S}_k^{J\pm}$.
\item 
The spinorial module is formed by the linear combination of vectors
  belonging to the half-integer spin spaces, $\Phi_{J,k+\half, m}^j$
  which can be obtained by the repeated action of $J_+$ (using the
  co-product) on
\bea
\Phi_{J,k+\half, -j}^{j}={\cal N}
a_{2}^{\dagger(j+k+\half)}a_{1}^{(j-k-\half)}.
\label{lowest}
\eea
This spinor module can be splitted into positive chiral and negative
chiral subspaces using the above $\Gamma$. Thus we have
\begin{equation}
    \label{eq:spinorial-module-splitted}
    \mathcal{S}_k^{J}=\mathcal{S}_k^{J \ +}\oplus\mathcal{S}_k^{J \ -}.
  \end{equation}
\end{itemize}
Now the Dirac operator which maps the spinor module to itself and also
anti-commutes with $\Gamma$ is defined using two operators $K_\pm$ and
acts on $\Psi$ as
\begin{equation}
  \label{eq:Dirac-operator}
  D\Psi=K_+\Psi+K_-\Psi.
\end{equation}
Here the $K_{\pm}$ are operators mapping $\mathcal{S}_k^{J \
  \pm}$ to $\mathcal{S}_k^{J \ \mp}$. In terms of the operators
$a_i^{\dagger},a_i,b^{\dagger},b$ we write the action of $K_{\pm}$ as
\begin{eqnarray}
  \label{eq:K-operators}
  K_+\Psi&=&ba_2^{\dagger}\Psi a_1^{\dagger}b-ba_1^{\dagger}\Psi a_2^{\dagger}b
  \\
   K_-\Psi&=&b^{\dagger}a_1\Psi a_2b^{\dagger}-b^{\dagger}a_2\Psi a_1b^{\dagger}.
\end{eqnarray}

We can easily check that with the above definitions, the
eigenfunctions of the Dirac operator are
$\Psi_{J,k,m}^{j\pm}=\fr{1}{\sqrt{2}}\left [\Phi_{J,k+\half,m}b \pm
  \Phi_{J,k+\half,m}b^\dagger\right]$ with eigenvalues
$\sqrt{(j+\half+k)(j+\half-k)}$. The $|M-N|$ zero modes are
\bea
&\Psi_{+0}^{{m_1}{m_2}}={\cal N}_1 a_{1}^{\dagger {m_1}}
a_{2}^{\dagger{m_2}}b^\dagger&\\
&\Psi_{-0}^{{n_1}{n_2}}={\cal N}_2 a_{1}^{n_1} a_{2}^{n_2}b&
\eea
where ${\cal N}_\a, \a=1, 2$ are proportionality constants and allowed
value of topological index $2k$ for the above zero modes are
$m_1+m_2+\mu>0$ and $n_1+n_2+\nu<0$ respectively and for both these zero
modes, the spin $j=|k|-\half$. Thus it is clear that the number of
zero modes are $2j+1=2|k|$ which is equal to $|M-N|$\cite{gkp}. It is
clear that the trace of $\Gamma$ restricted to the space of these zero
modes of the Dirac operator will count the net chiral zero modes
since each of the $\pm$ chiral zero modes will have $\pm 1$ as
eigenvalues under the chiral operator. We have seen that this number
is equal to $2k \in {\mathbb Z}$ which is the topological index of the
spinor field by construction. Thus we see here that the index theorem
is satisfied by the Dirac operator defined above on fuzzy sphere.

$D$ and $\Gamma$ as operators on the $U(su(2))$-module
$\mathcal{S}^J_k$ fulfill the conditions for the index theorem stated
above and they are invariant with respect to $su(2)$. The $su(2)$
invariant trace is given by the usual trace on the module
$\mathcal{S}_k^J$ and it is this trace we use in evaluating the index
of the Dirac operator, $D$.

\section{$\uq$ Invariant Dirac Operator $D_q$}

In this section we present an $\uq$ invariant Dirac operator $D_q$
defined on q-deformed fuzzy sphere. The $D_q$ and its spectrum were obtained in \cite{us}.
One of the interesting aspects of the spectrum of this Dirac
operator is the novel double degeneracy for the case where $q$ is root of unity. Also in this
case, we showed that there is a natural cut-off introduced by the root of
unity. We now briefly recall the main points of the derivation of $D_q$ which are 
necessary for the derivation of the index theorem.

\subsubsection*{Tensor Representations of $\uq$}

A tensor representation of a Hopf algebra is a representation of this
algebra on a tensor product of two vector spaces through the use of the
co-product. Therefore, in the present case, if $\Hil_l$ and $\Hil^*_l$ are
an $\uq$ irreducible representation and its dual, then the tensor
representation on the space $\Hil_l\otimes \Hil_l^*$ is given by
\begin{equation}
  \label{eq:tensor-rep-definition}
  \rho_{\otimes}(a)\equiv\rho\otimes\bar{\rho}(\Delta(a))\Hil_l\otimes \Hil_l^*\subset\Hil_l\otimes \Hil_l^*
\end{equation}
for $a\in\uq$ and $\rho:\uq\to Aut(\Hil_l)$ is a homomorphism of the algebra $\uq$ to
the algebra of automorphisms of $\Hil_l$ and $\bar{\rho}:\uq\to
Aut(\Hil^*_l)$ is also an homorphism of algebra. If $\Hil_l$ is the
irreducible representation of dimension $2l+1$, then we see immediately
that $\Hil_l\otimes\Hil^*_l$ is not irreducible. However, if $q$ is real
and positive it can be always decomposed into a direct sum of irreducible representations, i.e.,
\begin{equation}
  \label{eq:q-real-decomposition}
  \Hil_l\otimes \Hil_l^*=\bigoplus^{2l}_{j=0} \Hil_j
\end{equation}
where $\Hil_j$ are the irreducible representations of
dimension $2j+1$. Note that with additional conditions, such a
decomposition is possible for $q$ 
being root of unity also.

So, if $q=e^{\frac{2 \pi i}{p}}$, we restrict ourselves to irreducible
nilpotent (or classical) representations\footnote{If $q$ is root of
  unity there are two types of representations: i) nilpotent, which can
  be related to $su(2)$ irreducible representations; ii) cyclical, which
is a new type of representation with no analog to $su(2)$.}, in the sense that the
eigenvalues of $(J_+)^p,(J_-)^p$ are both zero, then the decomposition into
irreducibles \cite{gaume} is
\begin{equation}
  \label{eq:q-root-decomposition}
  \Hil_l\otimes \Hil_l^*=\bigoplus^{[2l,p-1]}_{j=0} \Hil_j,
\end{equation}
where $[2l,p-1]$ is $2l$ if $2l\leq p-1$ and
$\frac{p-1}{2}$ otherwise.

The tensor representation has for basis the set
$\{T_{jk}\}_{j=0,k=-j}^{2l,j}$. We also denote the angular momentum
quantum number $j$ as the rank of the tensor. The action of the algebra on
these tensors and their eigenvalues are given by
\begin{eqnarray*}
  \label{eq:tensor-rep-with-eigenvalues}
  \rho_{\otimes}(J_{\pm})T_{jk}=J_{\pm}T_{jk}K^{-\frac{1}{2}}-
  K^{-\frac{1}{2}}T_{jk}K^{\frac{1}{2}}J_{\pm}K^{-\frac{1}{2}}&=&\sqrt{[j\pm
    k +1][j\mp 1]}~T_{jk\pm 1}, 
\\
  \rho_{\otimes}(K)T_{jk}=KT_{jk}K^{-1}&=&q^kT_{jk}.
\end{eqnarray*}
Observe that these tensors act as operators on $\Hil_l$.

We also define the dual tensor representation. These are representations
$\bar{\rho}_{\otimes}\equiv\bar{\rho}\otimes\bar{\bar{\rho}}$. We denote
the irreducible basis of this dual representation by $T^{\ddagger}_{jk}$.
The relation between $T^{\ddagger}_{jk}$ and $T_{jk}$ is given by
\begin{equation}
  \label{eq:t-double-dagger}
  T^{\ddagger}_{jk}=(-1)^{-k}q^{\frac{k}{2}}T_{j -k}.
\end{equation}

It is important to note that using these tensor operators
with rank $j=\half$, we can construct the tensors of higher ranks using
the $q$-Clebsch-Gordan coefficients
\cite{qgroups}( Therefore it is enough to show the results
explicitly for the
$q$-tensor operators with $j=\half$).

Let us denote by $A^{\dagger}_i, A_i$, with $i=1,2$, two sets of
$q$-bosonic creation and annihilation operators. These operators
satisfy the following relations
\begin{eqnarray}
  \label{eq:q-heisenber-algebra}
  A_iA_i^{\dagger}-q^{\frac{1}{2}}A_i^{\dagger}A_i&=&q^{\frac{-N_i}{2}} \\
  \lbrack N_i,A_i^{\dagger} \rbrack &=& A^{\dagger}_i \\
  \lbrack N_i,A_i \rbrack &=&-A_i,
\end{eqnarray}
where $A_i^{\dagger}A_i=[N_i]$ and $N_i$ is the number operator. The
$q$-bosonic operators and the number operator satisfying these relations are
known as $q$-Heisenberg algebra.

We can write the tensor operators with rank $j=\half $\cite{qgroups} in terms of this
$q$-bosonic operators
\begin{eqnarray}
  \label{eq:q-tensor-operators-j-half}
  \alpha_1^{\dagger}&\equiv& T_{\half \half} =
  A^{\dagger}_1q^{-\frac{N_2}{4}} \\
  \alpha_2^{\dagger}&\equiv &T_{\half -\half} = q^{\frac{N_1}{4}}A^{\dagger}_2
\end{eqnarray}
and the dual representation (\ref{eq:t-double-dagger}) of these tensors as
\begin{eqnarray}
  \label{eq:q-tensor-operators-j-half-dual}
  \alpha_1&\equiv& T^{\ddagger}_{\half -\half} =
  A_1q^{-\frac{N_2+1}{4}} \\
  \alpha_2&\equiv &T^{\ddagger}_{\half \half} = - q^{\frac{N_1+1}{4}}A_2 .
\end{eqnarray}
The set of operators $\alpha_i^{\dagger},\alpha_i$ satisfies
\begin{eqnarray}
  \label{eq:relations-alpha}
  \alpha^{\dagger}_1\alpha^{\dagger}_2&=&q^{-\frac{1}{2}}\alpha^{\dagger}_2\alpha^{\dagger}_1
  \\
  \alpha_1\alpha_2&=&q^{\frac{1}{2}}\alpha_2\alpha_1 \\
  \alpha^{\dagger}_1\alpha_2&=&\alpha_2\alpha^{\dagger}_1 \\
  \alpha_1\alpha^{\dagger}_2&=&\alpha^{\dagger}_2\alpha_1.
\end{eqnarray}

\subsubsection*{Spinor Module}

The spinor module $\mathcal{S}$ is made of half-integer angular momentum tensor
operators combined with a set of fermionic creation and annihilation
operators $b^{\dagger},b$. This set of fermionic operators is the usual
ones, i.e., $\{b,b^{\dagger}\}=1$. A basis of this spinorial module is a monomial of the form
\begin{equation}
  \label{eq:q-tensor-operator-as-monomials}
  \a_1^{\dagger \ m_1}\a_2^{\dagger \ m_2}b^{\dagger \ \mu}\a_1^{n_1}\a_1^{n_2}b^{\nu}
\propto A_1^{\dagger \ m_1}A_2^{\dagger \ m_2}b^{\dagger \ \mu}A_1^{n_1}A_1^{n_2}b^{\nu}
\end{equation}
where the proportionality constant would involve $q$ and $N_\a$. In
the above $m_1,m_2, n_1, n_2$ are non-negative integers and $\m=\n=0,1,
\m+\n=1$. Also $m_1+m_2+\mu\le M$ and $n_1+n_2+\nu\le N$, $M-N=2k$,
$m_1+m_2 + \m-n_1-n_2-\n=2k$.  We also have $M+N=2J$ which act as the
fuzzy cut-off parameter. Similar to the previous section, we define
the spinorial fields as
\begin{equation}
  \label{eq:q-spinorial-field}
   \Psi=\Psi^+(A^{\dagger}_i,A_i,b^{\dagger},b)+\Psi^-(A^{\dagger}_i,A_i,b^{\dagger},b)=f(A^{\dagger}_i,A_i)b+g(A^{\dagger}_i,A_i)b^{\dagger},
\end{equation}
 such that
\begin{equation}
  \label{eq:q-spinorial-field-bosonic-part}
  \Psi^{\pm}=\sum_{m_1,m_2,\mu,n_1,n_2,\nu} c_{m_1,m_2,\mu,n_1,n_2,\nu}A_1^{\dagger \ m_1}A_2^{\dagger \
    m_2}b^{\dagger \ \mu}A^{n_1}_1A^{n_2}_2b^{\nu},
\end{equation}
where $c_{m_1,m_2,n_1,n_2}$ are complex numbers multiplied by factors
involving $q$ and $N_\a$. The bosonic part of $\Psi$ in
Eqn. (\ref{eq:q-spinorial-field}), i.e., $f$ and $g$ 
can be expressed in terms of the spin-half tensors belonging to
$\frac{M}{2}\otimes \frac{(N-1)}{2}$ and
$\frac{(M-1)}{2}\otimes\frac{N}{2}$ respectively and can be decomposed
into the direct sum of IRR of half-integer spins as
shown in Eqns. (\ref{dsum1}) and (\ref{dsum2}). The spinorial field $\Psi$ are
build by taking the linear combinations of the operators belonging to
these spin-half spaces $\Phi_{J,{k+\half},m}^j$ where the lowest weight
state is given by
\be
\Phi_{J,{k+\half}-j}^j=(A_{2}^{\dagger} q^{\fr{N_1}{4}})^{(j+k)}
(A_{1}q^{-\fr{N_2+1}{4}})^{(j-k)} .
\label{qlowest}
\ee

\subsubsection*{$q$-Deformed Dirac Operator}

The chirality operator is defined as in the usual $su(2)$ case
\begin{equation}
  \label{qchiralop}
  \Gamma_q\Psi\equiv\Gamma\Psi=-[b^{\dagger}b,\Psi],
\end{equation}
where the subscript $q$ is used to distinguish it from the chirality
operator on fuzzy sphere and  for comparison of results in the
$S_{qF}^2$ case to the usual fuzzy one. This chirality operator splits
the $\uq$ spinor module into $\pm$-chiral subspaces.

As in the usual $su(2)$ case we construct auxiliary operators $K_{\pm}$,
which map $\pm$ chiral subspace to $\mp$ chiral subspace. We require also
that these operators be invariant with respect to $\uq$. In
\cite{us}, it was shown that the operators that fulfill these
conditions are
\begin{eqnarray}
  \label{eq:q-K+-operator}
  K_+\Psi&=& q^{-\frac{k-m}{4}}K^{-\frac{1}{2}}b\left[A_1^{\dagger}\Psi
    A_2^{\dagger}q^{\frac{k}{2}}-A_2^{\dagger}\Psi A_1^{\dagger} \right]b\\
  K_- \Psi&=& q^{-\frac{k-m}{4}}b^{\dagger}\left[A_1\Psi
    A_2q^{\frac{k}{2}}-A_2\Psi A_1 \right]b^{\dagger}K^{-\frac{1}{2}}.
\end{eqnarray}
The $q$-deformed fuzzy Dirac operator $D_q$ is defined by
\begin{equation}
  \label{eq:q-dirac-operator}
  D_q\Psi=K_+\Psi+K_-\Psi
\end{equation}
and thus $D_q$ anti-commutes with the chirality operator $\Gamma_q$.
It is easy to see that the zero modes of this $q$-Dirac operator are
given by \cite{us}
\begin{eqnarray}
  \label{eq:q-zero-modes}
  \Psi_{0 +}^{\ m_1 m_2}&=&{\cal N}_1 A_1^{\dagger \ m_1}A_2^{\dagger \
    m_2}b^{\dagger} \\
  \Psi_{0 -}^{\ n_1 n_2}&=&{\cal N}_2 A_1^{n_1}A_2^{n_2}b.
\end{eqnarray}
where ${\cal N}_\a,~\textrm{with}~\a=1, 2$ are normalisation constants and they
involve factors of $q$ and the number operators $N_\a$. These zero
modes correspond to the angular momentum number
$j=|k|-\frac{1}{2}$. The number of these zero modes are given by $|M-N|=2j+1=2|k|$.
The non-zero eigenvalues of this Dirac operator are
$\sqrt{[j+\half+k][j+\half-k]}$ and corresponding eigenfunctions are
\be
\Psi_{J,k,m}^{j\pm}=\fr{1}{\sqrt{2}}\left [\Phi_{J,k+\half,m}b \pm
  \Phi_{J,k+\half,m}b^\dagger\right]
\ee
where $\Phi_{J,k\pm\half,m}$ is obtained from Eqn.(\ref{qlowest}). Here
we note that the eigenvalues are real. This is true even for
the case of $q=e^{\frac{2\pi i}{p}}$ provided $2l+1<p$ which is a constraint
one naturally introduce on the fuzzy cut-off parameter\cite{us}. This
condition is also needed to ensure the positivity of the inner product
as explained in the next section. The reality of the eigenvalues
clearly shows that our Dirac operator is self-adjoint as
required. We also note that the chirality operator $\Gamma$
anti-commutes with the Dirac operator and it splits the spinor module
into $\pm$-chiral subspaces.

\section{Index Theorem for $D_q$}

In the previous section we have presented the $\uq$ invariant Dirac operator which
anti-commutes with a chiral operator $\Gamma_q\equiv\Gamma$ and we have given
its zero modes\cite{us}. Both $D_q$ and $\Gamma_q$
act on an $\uq$ spinorial module $\mathcal{S}$. In order to
apply the index theorem to these operators we have to first define an $\uq$
invariant trace which is also essential in showing that $D_q$ is self-adjoint.

\subsubsection*{$\uq$-invariant Trace}

A trace acting on operators of a Hopf algebra module $V$ is said to be
invariant if it satisfies
\begin{equation}
  \label{eq:trace-invariance}
  Tr(\rho_{\otimes}(\Delta(a))\hat{T})=\epsilon(a)Tr(\hat{T}),
\end{equation}
where $\epsilon(a)$ is the co-unit of the element $a$ of the Hopf algebra
and $\hat{T}$ is an element of vector space $V\otimes V^*$. 

We can easily see that since the co-product of $\uq$ is deformed, the
above condition for the usual trace on the $\uq$-module $\Hil_l$ is not
satisfied. In order to have an invariant trace on a
representation $\Hil_l$, we need to deform the trace as well. The general
recipe to construct this deformed trace, which is called $q$-trace here,
is given in \cite{chari}. There it is shown that the usual trace is a
linear functional on the space $\Hil_l\otimes\Hil_l^*$. However, in general,
for quasi-triangular Hopf algebras this trace is not invariant. So a new
trace is defined to be a linear functional on the space
$\Hil^{**}_l\otimes\Hil_l^*$. It can be shown that
$\Hil^{**}_l\simeq\Hil_l$ as a representation for these quasi-triangular
Hopf algebras.

In the present case of $\uq$, the equivalent map $\Hil_l\otimes\Hil^*_l$ to
$\Hil^{**}_l\otimes\Hil^*_l$ is given by multiplication by the matrix representation of
the generator $K=q^{J_3}$, i.e., the tensor basis $T_{lm}\in\Hil_l\otimes\Hil^*_l$ goes to $K\cdot T_{lm}\in\Hil^{**}_l\otimes\Hil^*_l$. Thus, we define the $q$-trace, denoted by $Tr_q$, as
\begin{equation}
  \label{eq:q-trace-definition}
  Tr_q(T_{lm})=Tr(KT_{lm})
\end{equation}
where in the right hand side we have the usual matrix trace\cite{kullish}. Now, one can
easily check the invariance of this trace with respect to
$U_q(su(2))$. Indeed
\begin{equation*}
  Tr(\bar{\bar{\rho}}\otimes\bar{\rho}(\Delta(a))(KT_{lm}))=\epsilon(a)Tr(KT_{lm}).
\end{equation*}
Observe that we have to consider the proper representation with double bar
in the first term of the tensor product. We also note that
$\bar{\bar{\rho}}(a)=\rho(S^2(a))$, $S$ being the antipode.

In the space $\Hil_l\otimes\Hil^*_l$ we can define a positive
definite inner product using the $q$-trace as
\begin{equation}
  \label{eq:q-inner-product}
  \left(A,B\right)=Tr_q(A^{\ddagger}B),
\end{equation}
where $A,B\in\Hil_l\otimes\Hil^*_l$ and $A^{\ddagger}$ is in the dual of
$\Hil_l\otimes\Hil^*_l$.

It can be easily seen that the reality condition
$\overline{(A,B)}=(B,A)$ is satisfied for the case
of $q$ being real. But for $q$ being root of unity, it is 
 satisfied  only if we define the inner product with 
\begin{equation}
  \label{eq:quasi-star}
  A^{\circledast}=K^{-1}A^{\ddagger},
\end{equation}
such that $A^{\circledast \circledast}\neq A$ and
\begin{equation}
  \label{eq:q-inner-product-root}
  \left(A,B\right)=Tr_q(A^{\circledast}B)=Tr(A^{\ddagger}B),
\end{equation}
where on the right hand side $Tr$ is the usual trace. Thus we see that
the reality condition is satisfied for $q$ being root of unity also.
Further we observe that the positivity of this inner product is only
satisfied if $2l+1<p$, where $q=e^{\frac{2 \pi i}{p}}$. 

Thus we have now defined $\uq$-invariant trace which allows one to
define the self-adjoint of a given operator and are in a position
to evaluate the trace of the chirality operator on $S_{qF}^2$ and
obtain the index theorem.

\subsubsection*{Index Theorem for $D_q$}

The $\uq$ invariant self-adjoint Dirac operator $D_q$ and the chiral
operator $\Gamma_q$ on q-deformed fuzzy sphere do satisfy all the
requirements listed for a generic self-adjoint operator in section 2. 
We have also presented a $q$-trace (\ref{eq:q-trace-definition}) and the definition of
self-adjointness with respect with this new trace. Now, we are in position
to apply the index theorem to the operator $D_q$. We first calculate
the $q$-trace of $\Gamma$ and then the $q$-trace of a deformed opertator
$\tilde\Gamma$, which we will define.
\begin{enumerate}
\item{ Invariant trace of Chiral operator $(\Gamma_q)$ and Index} \\ \\
First we consider chirality operator defined on the q-deformed fuzzy
sphere given in Eqn. (\ref{qchiralop}) to derive the index. Using the
definition of the invariant trace and zero modes we get 
\begin{equation}
  \label{eq:q-index-theorem}
  Tr_{q,D=0}(\Gamma_q)=[n_+]-[n_-].
\end{equation}
This can be re-expressed using identities involving the $q$-numbers
\cite{qgroups} as
\begin{equation}
  \label{eq:q-index-theorem-2}
  [n_+]-[n_-]=\frac{[n_++n_-][\frac{(n_+-n_-)}{2}]}{[\frac{(n_++n_-)}{2}]}.
\end{equation}
Since $n_+-n_-$ is the number of net chiral zero modes of $D_q$ which
is equal to $2k$ and $n_++n_-=2J$ ($2J=M+N$ is the fuzzy cut-off paremeter), we get
\begin{equation}
  \label{eq:q-index-theorem-3}
  Tr_{q,D=0}(\Gamma_q)=\frac{[2J][k]}{[J]}.
\end{equation}
Observe that when $q\to 1$, we get $Tr(\Gamma)=2k$ which is the
correct result in this limit. Thus we observe that unlike in the case of usual fuzzy
sphere, here the trace of $\Gamma_q$ gives $q$-number of the
topological charge multiplied by factors which depend on the sum of
$n_{\pm}$. Thus we note that the trace of $\Gamma_q$ do not give just the
count of net chiral zero modes in the case of $S_{qF}^2$.

The $q$-trace of the identity operator on a vector space is also known as
the $q$-dimension or quantum dimension \cite{chari, gaume} of this vector
space. This quantity plays an important role in the representation theory
of quantum groups\cite{gaume}. Therefore, we can see the left hand side of
Eqn. (\ref{eq:q-index-theorem-3}) as the difference of the $q$-dimensions
of the subspaces of positive and negative chiralities of the spinor module
$\mathcal{S}$, respectively. The right hand side is a quantity depending
on the topological number $k$ of the spinorial field. We can thus consider
a generalization of the definition (\ref{index-dimension}) as
  \begin{equation}
    \label{eq:qindex-qdimension}
    qIndex ~D \equiv qdim (Ker D) - qdim (Ker D^\dagger) 
  \end{equation}
  where $qdim(Ker)$ is the $q$-dimension of the space of zero modes of the
  respective operators. Then, Eqn. (\ref{eq:q-index-theorem-3}) can be
  written as
  \begin{equation}
    \label{eq:qindex-2}
    qIndex~D=\frac{[2J][k]}{[J]}.
  \end{equation}
Here we note that the q-index depends on the fuzzy cut-off parameter $2J=M+N$. This novel feature of the q-deformed fuzzy sphere is absent in the usual fuzzy sphere where the fuzzy cut-off is not known to have any effect on the topological properties.

\item{Invariant trace of deformed Chiral operator $(\tilde\Gamma)$ and Index}\\ \\
Now we consider a deformed chirality operator ${\tilde\Gamma}=K^{-1}\Gamma$.
Since $DK=KD$ we have $DK^{-1}\Gamma+K^{-1}\Gamma D=0$. Though this 
deformed chiral operator $K^{-1}\Gamma$ is not an involution (
$K^{-1}\Gamma\ne 1$), we note that it also splits the spinor module
into $\pm$ chiral subspaces and in the limit $q\to 1$, it correctly
reduces to the chirality operator on $S_{F}^2$ as required. Due to the
$K^{-1}$ factor, when acted on the chiral spinors, it picks up and
extra $q^{-m}$ factor, apart from the $\pm 1$. With this deformed chirality operator we find
that
\begin{equation}
  \label{eq:q-chirality-op-index-theorem}
  Tr_{q,D=0}(K^{-1}\Gamma)=n_+-n_-=2k.
\end{equation}
Thus we see that the trace of the deformed chiral operator do give the
net number of chiral zero modes as in the usual case. We also know
that by construction\cite{us} this difference in the number of $\pm$
chiral zero modes is equal to the topological index $2k$ of the
spinorial field. Since here the trace have no dependency on fuzzy
cut-off or $q$, we are guaranteed to get the correct result in the
limit of $S_{F}^2$ and also in the limit of continuum sphere.
\end{enumerate}

Thus we see that the trace of chirality operator $\Gamma$ as well as
that of deformed chiral operator ${\tilde\Gamma}$ do reproduce the
expected result in the limit of $q\to 1$. But for generic $q$,
it is $\tilde\Gamma$ which gives the count of net chiral zero modes.
Thus it seems that the deformed chiral operator may be more natural
and of more use in
the study of chiral anomaly on deformed fuzzy sphere.

\section{Conclusion}

In this paper we have obtained the index of the Dirac operator defined on
q-deformed fuzzy sphere $S_{qF}^2$. Since the index is a measure of the
net chiral zero modes, we calculate it by evaluating the trace of the
chirality operator defined on $S_{qF}^2$. For this calculation we have used
the $\uq$ invariant trace\cite{kullish, chari}. We have shown that the
trace of chiral operator $\Gamma$ is proportional to the
$[n_+-n_-]_q$. The proportionality constant depends on the q-number of the 
fuzzy cut-off parameter. This situation is strickingly different from the undeformed fuzzy sphere where the cut-off parameter do not affect the topological invariants of the manifold. Using the explicit form of zero modes of the Dirac operator
$D_q$ on $S_{qF}^2$, we exhibit that $n_+-n_-=2k$ where $2k\in{\mathbb Z}$
is the topological index of the spinor field. Thus, with this chirality
operator, we get the index of Dirac operator proportional to the q-number
of the topological index. Since the $\uq$ invariant trace of the chirality
operator is the $q$-dimension of the eigenspace of $D_q$, this shows that
the $q$-dimension of the eigenspace is related to the topological index of
the spinor field.  Since the q-number reduces to usual number in the limit
$q\to 1$, we see that $[\frac{n_+-n_-}{2}]_q\to k$ in this limit as required. Here
we have seen that the proportionality constant depend on the sum of total
chiral zero modes and in the limit this proportionality constant becomes
unity. Then we have shown that a deformed chirality operator can be used
in place of $\Gamma$ and its $\uq$ invariant trace is just $2k$ as in the
usual case. Though this deformed chiral operator $\tilde\Gamma$ is not an
involutive operator, it does split the spinor module into $\pm$ chiral
subspace and also reduces to the correct chiral operator in the $q\to 1$
limit.

Using the $\uq$ invariant trace, we can construct the invariant
spinorial action on q-deformed fuzzy sphere $S_{qF}^2$. For this, we can use
$q$-Clebsch-Gordan techniques and combine $\Psi$, $D_{q}\Psi$ and
their duals defined using the invariant trace. Therefore terms like
$\Psi^{\ddagger}\Psi$ and $\Psi^{\ddagger}D_q\Psi$ are invariants with
respect to $\uq$ (For showing these we use Eqn.
(\ref{eq:t-double-dagger})).
Thus the $\uq$ invariant spinorial action is finally given by
\begin{equation}
  \label{eq:uq-spinorial-action}
  S=\frac{2\pi R^2}{[N+1]}Tr_q\left(\bar\Psi D_q\Psi+V({\bar\Psi}\Psi)\right),
\end{equation}
where, ${\bar\Psi}={\Psi}^{\ddagger}$, see (\ref{eq:q-inner-product}), or
$\bar{\Psi}={\Psi}^{\circledast}$, see (\ref{eq:q-inner-product-root}),
depending on whether $q$ is real or root of unity, $[N+1]$ is the
normalisation factor, $R$ is the radius of the underlying sphere and
$V({\bar\Psi}\Psi)$ is a potential function on the $\uq$ invariant
${\bar\Psi}\Psi$. It can be easily seen that the above action is also
invariant under chiral transformations for the massless case. Now, one may
ask if the chiral symmetry of the action is still a symmetry of the
corresponding quantum theory.

Consider the chiral invariant action ${\cal S}=S|_{V=0}$ where the action
$S$ is given in Eqn. (\ref{eq:uq-spinorial-action}). The corresponding
partition function is
\begin{equation}
  \label{eq:partition-function}
  \mathcal{Z}=\int \mathcal{D}{\bar \Psi}\mathcal{D}\Psi~ e^{-{\cal S}}.
\end{equation}
It is not invariant with respect to the chiral
transformations\footnote{The chiral transformations effected by the
  deformed chiral operator ${\tilde\Gamma}=q^{-J_3}\Gamma$ is
  ${\tilde\Gamma}\cdot\Psi=\Delta(q^{-J_3})e^{i\a\Gamma}(\Psi)$ where
  $\Delta(J_3)$ acts on the bosonic part, i.e., $f$ and $g$ of $\Psi$
  and $\Gamma$ acts on the fermionic part as in Eqn.(\ref{chiral}).}
\be
\Gamma:\Psi\to e^{i\a\Gamma}\Psi,~~~\Gamma:{\bar\Psi}\to\Psi
e^{i\a\Gamma}
\label{chiral}
\ee
where $\a$ is a real parameter. As in the case of spinorial action in the commutative space we note
that the contributions from $\pm$ chiral non-zero modes to the
Jacobian of chiral transformation cancel each other. But the
contribution coming from the zero modes do add up to give
\bea
  \label{eq:transformed-spinorial-measure}
  \mathcal{D}{\bar\Psi}\mathcal{D}\Psi&\to&
  e^{|n_{+}-n_{-}|}\mathcal{D}{\bar\Psi}\mathcal{D}\Psi\no\\
&=&e^{2k}\mathcal{D}{\bar\Psi}\mathcal{D}\Psi
\eea
which breaks the chiral symmetry of the quantum theory. We note here
that the above contribution to the integration measure is same for the chiral
transformations generated by $\Gamma$ as well as ${\tilde\Gamma}$.  
Thus we see that by defining an effective action with a counter-term that cancels
the above contribution, one can cancel the anomaly and retain the
chiral invariance. Since this new term has to be invariant under 
$\uq$, we need to define this term using $\uq$ invariant trace. Also
since the extra phase factor is just $2k$ and not proportional to $[k]_q$ 
we see that the effective action depend on the deformed chirality
operator ${\tilde\Gamma}$ rather than the
chirality operator $\Gamma_q$. Thus we get the effective action to be
\begin{equation}
  \label{eq:effective-action}
  {\cal S}_{eff}={\cal S}-Tr_{q}({\bar \Psi} K^{-1}\Gamma \Psi).
\end{equation}
Therefore, the deformed chirality operator may be better suited for
the study of spinor fields and their actions on $S_{qF}^2$.

In the commutative spaces, the new term in the effective action in
Eqn. (\ref{eq:effective-action}) have been expressed in terms of the gauge
fields alone\cite{topicsing}. There have been some studies aiming in
the construction of gauge field theories on fuzzy
sphere\cite{hs}. The path integral evaluation of above term can
lead to better understanding of the construction of gauge field action
on q-deformed fuzzy sphere as well as on fuzzy sphere.

\vspace{1cm}
\nn{\bf ACKNOWLEDGMENTS}\\
We thank A. P. Balachandran for many useful discussions and
comments. ARQ thanks FAPESP for support through the grant 02/03247-2. EH thanks V. O. Rivelles  for hospitality
and support at the Instituto de F\'{\i}sica, Universidade de S\~{a}o Paulo, Brazil.

\end{document}